\documentclass[pre,aps,floats,superscriptaddress,floatfix]{revtex4}
\usepackage{amssymb,amsmath}
\usepackage{amsmath,amssymb}
\usepackage{graphicx}
\usepackage{psfrag}

\def\beq{\begin{equation}}
\def\eeq{\end{equation}}
\def\bea{\begin{eqnarray}}
\def\eea{\end{eqnarray}}

\begin{document}
\title{Active to absorbing state phase transition in an evolving population 
with mutation}
\author{Niladri Sarkar}\email{niladri2002in@gmail.com}
\affiliation{Condensed Matter Physics Division, Saha Institute of
Nuclear Physics, Calcutta 700064, India}

\date{\today}
\begin{abstract}
We study the active to absorbing phase transition (AAPT) in a simple 
two-component
model system for a species and its mutant.
We uncover the non trivial 
critical scaling
behaviour and weak dynamic scaling near the AAPT that shows the significance of 
mutation and highlights the connection of this model
with the well-known directed percolation  universality class. Our model 
should be a useful starting point to study how mutation may affect extinction 
or survival of a species.
\end{abstract}

\maketitle

\section{Introduction}

Active to absorbing state phase transition (AAPT) forms a paradigmatic example 
of nonequilibrium critical phenomena~\cite{hinrichsen,tauber2}. 
In models exhibiting AAPT \cite{grass,carlon,urna1},
 a species  can exist in both the active and absorbing states, such that once it is in 
 the latter state, the transition probability to the active state vanishes. Simplest 
 models that exhibit AAPT often belong to  the well-known directed percolation (DP)
universality class. Some popular
examples of systems showing DP universal scaling
behavior \cite{takeuchi,urna2} are the epidemic process with recovery 
or the {\em Gribov process} \cite{grass-sund} and
the predator prey cellular automation models \cite{tauber3,tauber4,tome,tauber5}. 
In predator prey models for example \cite{rakesh,lipowski},
the growth (birth) and decay (death) of particles or species competes and thus
there may be a finite density of the species in the steady state ("active state") or 
extinction of the species ("inactive/absorbing state").
 Under the DP hypothesis \cite{Janssen2},
a system with a single absorbing state undergoing AAPT, shows  critical 
behavior belonging to the DP universality class in the absence
of any special symmetry, long range interactions, conservation law or 
quenched disorder. Else, non-DP like critical behavior
cannot be ruled out. In some cases, many absorbing states in an AAPT have 
also been found; see, e.g., in Refs.~\cite{dickman,satorras}. 

Continuum descriptions of AAPT in models displaying DP universality are based on the Reggeon field theory
\cite{gribov1,gribov2,mollison}, which is a stochastic multiparticle
process used to describe the local
growth of populations near their extinction threshold in an uniform environment \cite{Janssen3,grass1}.
The parameters of the model depends on the embedding environment
which are taken as constants  and their fluctuations ignored. 
If the fluctuations of the environment are taken into consideration, then 
whether the DP hypothesis and the DP universality class survive remains a 
question of general interest. Studies dealing
with the effect of environmental fluctuations on a species undergoing AAPT has
been made in Refs. \cite{nil1,nil2,antonov1,antonov2}.
It is now believed that nontrivial environmental dynamics and its feedback on the
species undergoing an AAPT substantially alter the critical exponents at the AAPT
leading to new universal behavior different from the DP universality class.
For instance, by considering  the environment to follow its own fluctuating scale
invariant dynamics,  Refs.~\cite{nil1,nil2} generically found non-DP like critical
scaling at the AAPT, often associated with {\em weak dynamic scaling}, where the
species undergoing AAPT and the environment have unequal dynamic exponents.
Ref.~\cite{nil2} also finds feedback of the species on the environment to be
relevant in determining the ensuing universal behavior. These are in general
modeled by coupling a second auxiliary dynamical field having its own dynamics
with the species that undergoes an AAPT.

Mutation of microbes and bacterial colonies has been an active area of research 
for quite some time now. Mutation in an evolving and growing population of a species can 
significantly alter its long-time state. Mutation in microbial colonies are important to 
understanding how the microbial population differentiates along the growing front in time and 
gives rise to well defined domains of different colonies \cite{alberti,erwin-paper,korolev}. 
For instance, if the mutation rate that sets the rate of creation of the mutant 
of the original species is large, 
but the back-mutation is small, it is conceivable that the original species 
will eventually go extinct, leaving only the mutant population as active. On 
the other hand, if the mutation rate is small compared to back mutation, the 
original species should continue to thrive with a small population of the 
mutant floating around in an otherwise pure species dominated world. Thus, 
depending upon the relative magnitudes of the mutation and the 
back mutation rate, the original species may become extinct by undergoing an 
AAPT~\cite{erwin-paper,lav-thesis}. This has impotant consequences specially in the 
formation of cancer and tumour cells in tissues. If mutation gives rise to 
deleterious population, a proliferation of the mutants might result in cancerous growth 
in a healthy tissue \cite{merlo,vogel,lav1,lav2}. So to contain the deleterious mutation, 
one can theoretically argue that the back mutation rate into the original species 
should be larger than the forward mutation rate. Since survival of the 
original species depends upon the suppressing mutation or facilitating 
back-mutation, it is conceivable that a back-mutation that is triggered by the 
presence of the original species may serve as a simple model of defence 
mechanism against proliferation of the mutant population, e.g., mutant cells in 
a body. We incorporate this in a simple way in our model below.

In this article, we propose a two-species nonconserved reaction-diffusion
model that describes the competing population dynamics of species A and and its mutant B, 
where the mutant B is allowed to back mutate into the pure species A. 
We study the AAPT displayed by it. Our model is
distinguished by the feature that the density of the 
mutant species B does not
obey any conservation law in the active state of the model, as a
result of its interaction with the species A, unlike the models in
Refs.~\cite{nil1,nil2}. Our model is well-suited to study whether or not
the lack of conservation laws for the mutant B dynamics due
to its coupling with the pure species A undergoing AAPT affects the critical
scaling of the AAPT. Apart from that, a more practical motivation of our model 
is definitely the production of mutants during the growth of a bacterial colony 
\cite{shapiro,nelson,neulinger,erwin-paper}. 
Despite the simplicity of our model, we obtain
a set of interesting results. For example, we find one physically stable fixed point 
with different dynamic exponents corresponding to the two species respectively,
when the diffusion coefficient of the species undergoing AAPT
is much greater than the diffusion coefficient of the mutant species,
which in turn follows a conservation law in the absence of the pure species. This phenomena is
commonly referred to as weak dynamic scaling in phase transition literature.
We find that the 
fixed point exhibiting weak dynamic scaling shows DP like universal behaviour, with 
exactly the same critical exponents as the DP universality class, 
which we argue as purely coincidental and is a consequence of our one loop Dynamic
Renormalization Group (DRG) analysis. Strong dynamic scaling with same dynamic exponent
of the two fields are expected when the diffusion coefficient of the two species
are of the same order, a feature we have not discussed in this article.
The rest of the paper is organized as
follows: In Section. \ref{mod} we introduce our model
following a brief review of the DP universality class. In Section \ref{rg},
we do a dynamic renormalization of our model using the DRG procedure. In
Section \ref{fp}, we find out the fixed points and the corresponding
critical exponents in the weak dynamic scaling regime. In Section. \ref{con}
we conclude our study with a summary of our results.

\section{The model and equations of motion}
\label{mod}

In this Section we introduce our model of population dynamics of species  
A and its mutant 
B with densities $\rho({\bf x},t)$ and $\phi({\bf x},t)$
respectively.
Due to the mutation of A to B and the latter's back mutation to A lead to 
nonconservation of B in the active state of A.
 Before going to the details of
our model, we recall the DP model in brief.

\subsection{Directed Percolation model}
\label{review}

Consider a population dynamics model in which the growth of the population is
linearly dependent on the local species density given by $\phi({\bf x},t)$,
and the death is proportional to the square of the species density, which
describes death due to overcrowding. The species density undergoes a
nonequilibrium AAPT, whose long wavelength, large time behaviour is described
by the DP universality class. The Langevin equation for such a
population dynamics model can be written in terms of species density as
\cite{tauber1}
\begin{equation}
\frac{\partial\phi}{\partial t} = D\nabla^2\phi + \lambda_g\phi -
\lambda_d\phi^2 + \sqrt\phi \zeta,\label{dpeq}
\end{equation}
where the first term on the right hand side is the diffusion term with $D$
as the diffusion coefficient, $\lambda_g$ is the birth rate and
$\lambda_d$ is the rate of death due to overcrowding.
The stochastic function $\zeta({\bf x},t)$
is a Gaussian distributed white noise with zero mean and a variance
\begin{equation}
\langle\zeta({\bf x},t)\zeta(0,0)\rangle = 2D_2 \delta ({\bf
x})\delta (t). \label{vari1}
\end{equation}
The multiplicative nature of the effective noise ensures the existence
of an absorbing state ($\phi=0$). On dimensional ground, a
characteristic length scale $\xi \sim \sqrt{D/|\lambda_g|}$ and a diffusive
time scale $t_c\sim \xi^2/D\sim 1/|\lambda_g|$ can be derived from
Eq.~(\ref{dpeq}), with both diverging upon approaching the critical
point $\lambda_g=0$. The critical exponents may be defined in the usual way
\cite{tauber1}
\begin{equation}
\langle \phi ({\bf x},t\rightarrow\infty)\rangle \sim
\lambda_g^\beta,\;\;\langle \phi({\bf x},t)\rangle \sim
t^{-\alpha}\; (\lambda_g=0),\;\; \xi\sim \lambda_g^{-\nu},\;\;
t_c\sim \xi^z_\phi/D\sim \lambda_g^{-z_\phi\nu}, \label{scaling}
\end{equation}
with the mean field scaling exponents given by
\begin{equation}
\beta=1,\alpha=1,\nu=1/2,\,\, {\mbox {and}},\,\, z_\phi=2.\label{mean}
\end{equation}
The anomalous dimension $\eta$, which characterises the
scaling of the two-point correlation function, is zero
\cite{tauber1} in the mean field limit. To find out how the fluctuations
affect the mean field scaling exponents, a dynamic renormalization group (DRG)
calculational scheme is used to find out the corrections to the bare
correlation and vertex functions in the model. It should be noted that
the Janssen-de Dominicis action functional which corresponds to
the Langevin Eq.~(\ref{dpeq}), has an invariance under rapidity symmetry
given by $\hat\phi ({\bf x},t) \leftrightarrow
\phi({\bf x},-t)$~\cite{tauber1}, where $\hat\phi$ is the
auxilliary field conjugate to $\phi$~\cite{tauber1}.
Invariance under rapidity symmetry is a signature of the DP universality class
and all models whichever falls under the DP universality class should be
invariant under the rapidity symmetry asymptotically. By performing
a perturbative expansion in $\epsilon=d_c-d$, $d_c=4$ is the upper
critical dimension for this model using the DRG scheme, one obtains \cite{tauber1},
\begin{equation}
z=2-\epsilon/12,\eta=-\epsilon/12 \,\,\mbox{and}\,\, {1 \over \nu}=2
- \epsilon/4. \label{dpexpo}
\end{equation}
These universal scaling exponents characterise the DP universality class.
As the DP hypothesis \cite{Janssen2}, suggest that the DP universality
class is very robust, any one of the conditions of the DP hypothesis are
to be violated in order to find new scaling behaviour.
Refs.~\cite{antonov1,antonov2,nil1} have shown that fluctuating environments
with spatially long-ranged noises can modify the scaling behaviour of the DP
universality. In this article we have introduced a reaction-diffusion
model involving two species and studied how the interdependence
of the two species on their mutual birth and death affects the scaling
properties of the DP universality class.

\subsection{Two species reaction diffusion model}


Our two-species model consists of the species A and its mutant B. Species 
A 
reproduces at  a given rate; it can also mutate to species B and also 
die due to overcrowding at fixed rates. Naturally, proliferation of the mutant B, 
if unchecked, should lead to eventual extinction of A.  In order to enlarge the scope of our 
model, we allow back-mutation from species B to A, ensuring a compettition 
between the original species and the mutant. We consider the specific 
case where back-mutation of B to A is triggered by the presence of A locally. 
Thus, species B can back-mutate to A at a given rate, provided species A is 
available in its neighborhood. Our choice for the specific 
form of back mutation, though admittedly over-simplified, serves several purposes. 
For instance, since the back mutation is facilitated by the presence of A, 
it suggests that the original species has an ability to suppress effects of 
(unwanted or random) mutations, necessary for its survival as a species. 
In addition, it is consistent with an absorbing state transition of A with 
the system being filled up with B, which we are interested to study. 
Together with the other processes described above it provides a 
minimal model to study mutation and back-mutation in population 
dynamics of an evolving species and their effects on the AAPT in 
the model. The
 two Langevin equations for the densities $\rho$ and $\phi$:
\bea
{\partial \rho \over \partial t} &=& D_\rho\nabla^2\rho +(1-\lambda_1)\rho+ 
\lambda_g\rho\phi
-\lambda_d\rho^2 +\sqrt{\rho}\eta,
\label{rho} \\
{\partial \phi \over \partial t} &=& D_\phi\nabla^2\phi + \lambda_1\rho
-\lambda_2\rho\phi +\sqrt{\rho}\xi. \label{phi}
\eea
In Eq.~(\ref{rho}), the first term on the right hand side represents 
diffusion
of species A with a diffusion coefficient $D_\rho$.  The second term 
with $1-\lambda_1>0$ represents
growth (reproduction) of A at rate $1-\lambda_1$. The third term
represents the growth in population of A due to back-mutation of B with a rate  
$\lambda_g\phi,\,\lambda_g>0$.
 The
next term is a decay term ($\lambda_d>0$) which represents the death of $\rho$ due to
overcrowding. The
stochastic noise $\eta({\bf x},t)$ is the Gaussian distributed white noise
with zero mean and a variance
\bea
\langle \eta({\bf x},t)\eta(0,0)\rangle=2D_2\delta({\bf x})\delta(t). \label{eta}
\eea
The multiplicative nature of the noise in Eq.~(\ref{rho}) ensures the existence of an absorbing
state ($\rho=0$).
The dynamics of species B, as given by Eq.~(\ref{phi}), is a combination of
diffusion with diffusion coefficient  $D_\phi$, production of B through
mutation of A to B at rate $\lambda_1$ and back-mutation of B by A at
rate $\lambda_2\rho\phi$, $\lambda_1,\lambda_2>0$. Clearly, back-mutation of 
B can take place only if there are some species A around locally. We assume 
for simplicity
that the only source of stochasticity in the dynamics of $\phi$ is $\rho$,
and hence, we model it by a multiplicative noise $\sqrt\rho\xi$, such that
in the absorbing state, the dynamics of B is noise-free.
We choose $\xi$ to be a Gaussian distributed
white noise with zero mean and a variance
\bea
\langle \xi({\bf x},t)\xi(0,0)\rangle=2D_1\delta({\bf x})\delta(t). \label{xi}
\eea
Evidently, our model as given by Eqs.~(\ref{rho}) and (\ref{phi}) admits $\rho=0,\,
\phi=const.\neq 0$ as an absorbing state. One may also add a conserving additive noise in Eq.~(\ref{phi}),
reflecting the thermal fluctuations of $\phi$. This noise would then have
survived in the absorbing state. We neglect this noise for simplicity, which
is akin to assume a "low temperature limit" for species B. Interestingly, in the
absence of an additive conserved noise in (\ref{phi}), $\rho=0,\,\phi=0$ is also
an absorbing state. We ignore this and focus on the absorbing state 
$\rho=0,\,\phi=\phi_0=const.\neq 0$.
It is instructive compare Eqs.~(\ref{rho}) and (\ref{phi}) with the model 
in Ref.~\cite{nil2}. In Ref.~\cite{nil2}, the second field is a conserved field 
in both the active and absorbing states of the species, and hence is 
inappropriate to model a mutant. In contrast, $\phi$ here is {\em 
non-conserved} in the active state of A, appropriate to model population 
changes of the mutant due to mutation or back-mutation. It is only in the 
absorbing state of A that $\phi$ is conserved; see also 
Ref.~\cite{erwin-paper}. This feature clearly distinguishes our model from 
Ref.~\cite{nil2}.

We write
$\phi=\phi_0 +\delta\phi$. This modifies (\ref{rho}) and
(\ref{phi}) to
\bea
{\partial \rho \over \partial t} &=& D_\rho\nabla^2\rho +r\rho +
\lambda_g\rho\delta\phi -\lambda_d\rho^2 +\sqrt{\rho}\eta,
\label{rhofluc} \\
{\partial \delta\phi \over \partial t} &=& D_\phi\nabla^2\delta\phi + \lambda_3\rho
-\lambda_2\rho\delta\phi +\sqrt{\rho}\xi, \label{phifluc}
\eea
where $r=1-\lambda_1+\lambda_g\phi_0$ and 
$\lambda_3=\lambda_1-\lambda_2\phi_0$. 
Coupling constant $\lambda_3$ should be positive so as to prevent $\phi$ from collapsing into an absorbing
state in the presence of $\rho$, without passing through an active configuration.
Now denoting $\delta\phi$ as $\phi$ so as to avoid notational complexity, the
equations of motion for the two fields in the model can be written as
\bea
{\partial \rho \over \partial t} &=& D_\rho\nabla^2\rho +r\rho +
\lambda_g\rho\phi -\lambda_d\rho^2 +\sqrt{\rho}\eta,
\label{rhofinal} \\
{\partial \phi \over \partial t} &=& D_\phi\nabla^2\phi + \lambda_3\rho
-\lambda_2\rho\phi +\sqrt{\rho}\xi. \label{phifinal}
\eea

We redefine the coefficients $r=D_\rho\tau$ and $\lambda_d=D_\rho g_1/2$ for
calculational convinience so that Eq.~(\ref{rhofinal}) now takes the form
\bea
{\partial \rho \over \partial t} = D_\rho(\tau +\nabla^2)\rho
+ \lambda_g\rho\phi -{D_\rho g_1 \over 2}\rho^2
+\sqrt{\rho}\eta. \label{rhored}
\eea
The critical point is given by renormalized $\tau=0$. To what extent the
nonlinear couplings in our model alter the mean field DP exponents given by (\ref{dpexpo})
may be answered systematically by using the standard one-loop
 Dynamic Renormalization Group (DRG) framework. This requires calculating
 the primitively divergent vertex functions in the model up to the one-loop
 order in expansions in terms of the effective coupling constants and
 absorbing the divergences in redefined or {\em renormalized} parameters
 of the model. These allow us to obtain the renormalized vertex or correlation
 functions in the model, from which the critical scaling exponents may be obtained.
 See \cite{zinn} for detailed technical discussions on the DRG technique.

Using the Langevin equations (\ref{rhored}) and (\ref{phifinal}), together with
the corresponding noise variances (\ref{eta}) and (\ref{xi}), the Janssen-De Dominics
\cite{de-dom} generating functional can be constructed which can be written as
\bea
\mathcal{Z}= \int D\rho D\hat\rho D\phi D\hat\phi \exp[-S], \label{gen}
\eea
where $\hat\rho$ and $\hat\phi$ are the auxilliary fields corresponding to the
dynamical fields $\rho$ and $\phi$ respectively, which enters the Eq.~(\ref{gen})
after elimination of the noises from the generating functional $\mathcal{Z}$.
For calculational convenience we redefine $i\hat\phi\rightarrow \hat\phi$,
$i\hat\rho\rightarrow \hat\rho$ and $D_2={D_\rho g_2 \over 2}$ in the generating
functional $\mathcal{Z}$.
The dynamical action functional $S$ corresponding to the model is then
given by
\bea
S &=&\int d^dx\int dt \hat\rho\{\partial_t
+D_\rho(-\tau +\nabla^2)\}\rho +
\int d^dx\int dt \hat\phi\{\partial_t + D_\phi \nabla^2\}
\phi - \lambda_3\int d^dx\int dt
\hat\phi\rho - \lambda_g\int d^dx\int dt
\hat\rho\rho\phi \nonumber \\
&&-D_1\int d^dx\int dt \hat\phi\hat\phi\rho +
\lambda_2\int d^dx\int dt \hat\phi\phi\rho -{D_\rho g_2 \over 2}
\int d^dx\int dt \hat\rho\hat\rho\rho + {D_\rho g_1 \over 2}
\int d^dx\int dt \hat\rho\rho\rho. \label{action}
\eea
Unlike the pure DP problem, in Eq.~(\ref{action}), the last two terms
have different coefficients $D_\rho g_1/2$
and $D_\rho g_2/2$, due to the breakdown of the rapidity symmetry by
the couplings $\lambda_g$, $\lambda_2$ and $D_1$.
 In addition, time can be rescaled to absorb $D_\phi$ in Eq.~(\ref{action}).

 In a na\"ive perturbative expansion, $\lambda_g,D_1,\lambda_2$ and $u\equiv g_1 g_2$
 appear as the expansion parameters. Rescaling space and time, it is straightforward
 to show that  the upper critical dimension $d_c=4$ for the coupling constants
 $\lambda_g,D_1,\lambda_2$ and $u$. We also introduce a control parameter
$\theta=D_\phi/D_\rho$ known commonly as the Schmidt number which determines
the ensuing nonequilibrium steady state (NESS) of our model. If the
renormalized versions of $\lambda_g,D_1,\lambda_2$ are non-zero at the DRG
fixed points (FP) at $d<d_c$, then new universal critical scaling behavior
is expected to emerge at the DRG FPs, such that the critical exponents would
pick up values different from their values for the DP universality class.
In the DRG analysis of our model, $\epsilon=d_c-d=4-d$ appears as a small
parameter; see Ref.~\cite{tauber} for the detailed technical discussions
on DRG applications in  the DP problem.

We begin by identifying the primitively divergent vertex
functions in model (\ref{action}). The vertex functions of our model are defined formally
by taking the appropriate functional derivatives of the vertex generating
functional $\Gamma[\rho,\hat\rho,\phi,\hat\phi]$ with respect to the various
fields $\rho$, $\hat\rho$, $\phi$ and $\hat\phi$,
with $\Gamma[\rho,\hat\rho,\phi,\hat\phi]$ being the Legendre transform
of $\log\mathcal{Z}$ \cite{zinn,tauber}:
 \bea
 \Gamma_{a_1a_2...a_n}\equiv \frac{\delta^n \Gamma}{\delta a_1 \delta a_2...\delta a_n},
 \eea
 where $a_1, a_2,...,a_n$ are the fields $\rho,\hat\rho,\phi,\hat\phi$.
 The bare vertex functions in our model that have divergent one-loop corrections
 are listed in Appendix~\ref{appen1}.

\section{Renormalization Group analysis}
\label{rg}

To renormalize the vertex functions we choose $\tau=\mu^2$ as the appropriate
normalization point about which the vertex corrections are found out up to
one-loop order, with $\mu$ being the intrinsic momentum scale of the renormalized
theory. Next we need the multiplicative renormalization $Z$-factors to determine
the scale dependence of the renormalized vertex functions on the momentum scale
$\mu$. This is possible as the $Z$-factors absorb the ultraviolet divergences
arising out of the one-loop integrals which makes the resulting theory finite.
The $Z$-factors for the various fields and parameters are defined as follows:
\bea
&&\phi=Z_\phi \phi^R \,,\, \rho=Z_\rho \rho^R\,,\,\hat \rho=Z_{\hat \rho}\hat
\rho^R\,,\,\hat\phi=Z_{\hat\phi}\hat\phi^R\,,\,D_\rho=Z_{D_\rho}D_{\rho}^R\,,\,
\lambda_g=Z_{\lambda_g}\lambda_g^R \,,\, g_1=Z_{g_1}g_1^R\,,\,
g_2=Z_{g_2}g_2^R\,,\,\tau=Z_\tau \tau^R,\nonumber \\
&&\lambda_3=Z_{\lambda_3}\lambda_3^R, \,
\lambda_2=Z_{\lambda_2}\lambda_2^R, \, D_\phi=Z_{D_\phi}D_\phi^R, \,\,
D_1=Z_{D_1}D_1^R, \label{zfacdef}
\eea
where the superscript $R$ refers to renormalized quantities. The various
$Z$-factors can be found out from the normalization conditions; see Appendix~\ref{appen2}.
Thus we have eleven renormalized vertex functions in comparison with the
thirteen $Z$-factors defined here. Therefore, there are two
redundant $Z$-factors which can be chosen arbitrarily. We hence use this freedom
to set $Z_\rho=Z_{\hat\rho}$ and $Z_\phi=Z_{\hat\phi}$. The $Z$-factors
calculated from the one-loop irreducible diagrams are found to be
\bea
Z_\rho &=& Z_{\hat\rho}=1+ {g_1g_2 \over 8}{\mu^{-\epsilon} \over 16\pi^2\epsilon},
\label{zrho} \\
Z_{D_\rho} &=& 1 - {g_1g_2 \over 8}{\mu^{-\epsilon} \over 16\pi^2\epsilon},
\label{zdrho} \\
Z_\tau &=& 1 + {3g_1g_2 \over 8}{\mu^{-\epsilon} \over 16\pi^2\epsilon},
\label{ztau} \\
Z_{g_1} &=& 1 + {3g_1g_2 \over 4}{\mu^{-\epsilon} \over 16\pi^2\epsilon}
+ {4\lambda_g^2D_1 \over D_\rho g_1D_\phi(D_\rho +D_\phi)}
{\mu^{-\epsilon} \over 16\pi^2\epsilon}, \label{zg1} \\
Z_{g_2} &=& 1+ {3g_1g_2 \over 4}{\mu^{-\epsilon} \over 16\pi^2\epsilon},
\label{zg2} \\
Z_{\lambda_2} &=& 1 - {g_1g_2 \over 8}{\mu^{-\epsilon} \over 16\pi^2\epsilon}
+ {\lambda_2 g_2 \over (D_\rho +D_\phi)}{\mu^{-\epsilon} \over 16\pi^2\epsilon},
\label{zlambda2} \\
Z_{\lambda_3} &=& Z_\phi^{-1}[1-{g_1g_2 \over 8}{\mu^{-\epsilon} \over 16\pi^2\epsilon}
- {\lambda_2 g_2 \over (D_\rho +D_\phi)}{\mu^{-\epsilon} \over 16\pi^2\epsilon}],
\label{zlambda3} \\
Z_{\lambda_g} &=& Z_\phi^{-1}[1+ {g_1g_2 \over 4}{\mu^{-\epsilon} \over 16\pi^2\epsilon}
+ {\lambda_2 g_2 \over (D_\rho +D_\phi)}{\mu^{-\epsilon} \over 16\pi^2\epsilon}].
\label{zlambdag}
\eea
We also find that there are no one loop corrections to $\Gamma_{\hat\phi\phi}$,
which means that $Z_\phi Z_{\hat\phi}=1$. Using the choice $Z_\phi=Z_{\hat\phi}$, we
get $Z_{\phi}=1=Z_{\hat\phi}$. Thus Eqs.~(\ref{zlambda3}) and (\ref{zlambdag}) have
only unity as contributions coming from $Z_{\phi}$
\bea
Z_{\lambda_3} &=& 1-{g_1g_2 \over 8}{\mu^{-\epsilon} \over 16\pi^2\epsilon}
- {\lambda_2 g_2 \over (D_\rho +D_\phi)}{\mu^{-\epsilon} \over 16\pi^2\epsilon},
\label{zlambda3mod} \\
Z_{\lambda_g} &=& 1+ {g_1g_2 \over 4}{\mu^{-\epsilon} \over 16\pi^2\epsilon}
+ {\lambda_2 g_2 \over (D_\rho +D_\phi)}{\mu^{-\epsilon} \over 16\pi^2\epsilon}.
\label{zlambdagmod}
\eea

For the purpose of calculational convenience, we define three 
dimensionless constants $\alpha,\gamma$ and $\psi$ through
\bea
\lambda_2=D_\rho g_1\alpha \, ,\, \lambda_g=D_\rho g_1\gamma \,,\, D_1=D_\rho 
g_2\psi.
\label{res}
\eea
In what follows below, we treat $\alpha,\gamma,\psi$ as the coupling 
constants in the present problem
without any loss of generality. From the physical interpretations of 
the different constants in the present model all of $\alpha,\gamma$ 
and $\psi$ should be positive. Clearly, if $\alpha=0=\gamma$, there 
is no 
back-mutation.
 Eq.~(\ref{res})
gives us the multiplicative $Z$-factors of $\alpha$, $\gamma$ and $\psi$ as
$Z_\alpha=Z_{\lambda_2}Z_{D_\rho}^{-1}Z_{g_1}^{-1}$, $Z_\gamma=Z_{\lambda_g}
Z_{D_\rho}^{-1}Z_{g_1}^{-1}$ and $Z_\psi=Z_{D_1}Z_{D_\rho}^{-1}Z_{g_2}^{-1}$.
Their explicit values in terms of the effective coupling constant 
\bea
u=g_1g_2, \label{udef}
\eea
and Schmidt number $\theta= D_\phi/D_\rho$ take the form
\bea
Z_\alpha &=& 1 - {3u\mu^{-\epsilon} \over 4\epsilon}+ {\alpha u\mu^{-\epsilon}
\over (1+\theta)\epsilon} - {4u\gamma^2\psi\mu^{-\epsilon} \over
\theta(1+\theta)\epsilon}, \label{zalpha} \\
Z_\gamma &=& 1 - {3u\mu^{-\epsilon} \over 8\epsilon} + {u\alpha\mu^{-\epsilon}
\over (1+\theta)\epsilon} - {4u\gamma^2\psi\mu^{-\epsilon} \over \theta(1+\theta)
\epsilon}, \label{zgamma}\\
Z_{\psi} &=& 1- {3u\mu^{-\epsilon} \over 4\epsilon} + {2u\alpha\mu^{-\epsilon}
\over (1+\theta)\epsilon}, \label{zpsi}
\eea
where we have absorbed a factor of $1/16\pi^2$ in the definition of $u$. With
$u=g_1g_2$, the multiplicative $Z$-factor for $u$ is given by $Z_u=Z_{g_1}Z_{g_2}$
and $Z_\theta=Z_{D_\phi}Z_{D_\rho}^{-1}$. As $Z_{D_\phi}=1$, due to lack of
renormalization of $\Gamma_{\hat\phi\phi}$, $Z_\theta=Z_{D_\rho}^{-1}=1+
{u\mu^{-\epsilon} \over 8\epsilon}$.
Thus, we have identified the effective coupling constants for the model
to be $u$, $\alpha$, $\gamma$ and $\psi$. From the $Z$-factors of these couplings,
the $\beta$-functions corresponding to the renormalized coupling constants $u^R$,
$\alpha^R$, $\gamma^R$ and $\psi^R$ can be obtained, given by
\bea
\beta_u &=& u^R[-\epsilon + {3u^R \over 2} + {4u^R(\gamma^{R})^2\psi^R \over
\theta^R(1+\theta^R)}], \label{betau} \\
\beta_\alpha &=& \alpha^R[-{3u^R \over 4}+{u^R\alpha^R \over (1+\theta^R)}-
{4u^R(\gamma^R)^2\psi^R \over \theta^R(1+\theta^R)}], \label{betaalpha} \\
\beta_\gamma &=& \gamma^R[-{3u^R \over 8} + {u^R\alpha^R \over (1+\theta^R)}
- {4u^R(\gamma^R)^2\psi^R \over \theta^R(1+\theta^R)}], \label{betagamma} \\
\beta_\psi &=& \psi^R[-{3u^R \over 4}+{2u^R\alpha^R \over (1+\theta^R)}].\\
\beta_\theta &=& \theta_R [\frac{u^R}{8}].\label{betatheta}
\label{betapsi}
\eea
The zeros of the $\beta$-functions gives us the fixed point (FP) solutions 
for the 
model, i.e., by setting the rhs of Eqs.~(\ref{betau}-\ref{betatheta}).  The FPs 
may be obtained in three different physical limits, {\em viz.}, 
$\theta_R\rightarrow 0,\infty$ and $\theta_R$ remaining finite. The first two 
cases should be characterized by {\em weak dynamic scaling}, 
i.e., by $z_\rho<z_\phi$ and $z_\rho >z_\phi$, respectively. In contrast, the 
third possibility corresponds to $z_\rho=z_\phi$, implying {\em strong dynamic 
scaling}. In the limit of $\theta_R\rightarrow\infty$, $\beta_\theta=0$ yields 
$u_R=0$, which though satisfies $\beta_u=0$ is not a stable FP; or, 
equivalently, even for a very small $u_R$, $\beta_\theta$ shoots up to 
$\infty$. We thus discard this possibility.
 To find the fixed points for a finite $\theta$ value, the full 
$\beta$-functions 
should
be equated to zero, which is a highly daunting task given the complexity of the
equations involved. But for $\theta^R\rightarrow 0$, 
the algebra is tractable. Hence we settle for the tractable $\theta^R\rightarrow 0$
limit here and leave out the case of finite $\theta^R$.

\section{Fixed point analysis and critical exponents}
\label{fp}

In this section we perform a fixed point analysis corresponding to the $\theta^R\rightarrow 0$ limit.
 In this limit $D_\rho^R\gg D_\phi^R$, so that a weak dynamic scaling with
$z_\rho <z_\phi$ is expected at the stable FP.
The FPs are given by the solutions of the equations
\bea
{3u^R \over 2} + {4u^R(\gamma^R)^2\psi^R \over \theta^R}=\epsilon, \label{ubetaeq}
\\
-{3u^R \over 4}+u^R\alpha^R-{4u^R(\gamma^R)^2\psi^R \over \theta^R}=0,
\label{alphabetaeq} \\
-{3u^R \over 8} + u^R\alpha^R - {4u^R(\gamma^R)^2\psi^R \over \theta^R}=0,
\label{gammabetaeq} \\
-{3u^R \over 4}+2u^R\alpha^R=0, \label{psibetaeq}
\eea
from which nontrivial fixed points corresponding to $u^R\neq 0$, $\alpha^R\neq 0$,
$\gamma^R\neq 0$ and $\psi^R\neq 0$ are obtained. As we can see
$u^R\sim O(\epsilon)$, but $\alpha^R$, $\gamma^R$ and $\psi^R$ are just
numbers. To extract physically meaningful fixed points from the
eqs.~(\ref{ubetaeq}), (\ref{alphabetaeq}), (\ref{gammabetaeq}) and
(\ref{psibetaeq}), the terms should be divergence free in the
$\theta^R\rightarrow 0$ limit. But as can be seen the equations (\ref{ubetaeq}
-\ref{psibetaeq}) contains $\theta^R$ in the denominator making them
diverge in the $\theta^R\rightarrow 0$ limit. To make them free from
divergences, one should scale $(\gamma^R)^2\psi^R\sim \theta$. Assuming this
scaling to hold good in our case, we take 
\bea
m=u\alpha, \label{mdef}
\eea 
and
\bea 
t={u\gamma^2\psi \over \theta}, \label{tdef}
\eea 
as the effective coupling constants in the
limit $\theta^R\rightarrow 0$. Evidently, both $m$ and $t$ should be 
positive on physical ground. The 
renormalized effective couplings are hence
$m^R=u^R\alpha^R$ and $t^R={u^R(\gamma^R)^2\psi^R \over \theta^R}$ and their
corresponding $Z$-factors given by $Z_m=Z_uZ_\alpha$ and $Z_t=Z_uZ_{\gamma}^2
Z_\psi Z_\theta^{-1}$ respectively. The explicit form of the $Z$-factors
for the effective coupling constants then turn out to be
\bea
Z_u &=& 1+ {3u\mu^{-\epsilon} \over 2\epsilon} + {4t\mu^{-\epsilon} \over
\epsilon}, \label{zu0fin} \\
Z_m &=& 1+ {3u\mu^{-\epsilon} \over 4\epsilon} +{m\mu^{-\epsilon}
\over \epsilon}, \label{zm0fin} \\
Z_t &=& 1- {u\mu^{-\epsilon} \over 8\epsilon} + {4m\mu^{-\epsilon}
\over \epsilon} - {4t\mu^{-\epsilon} \over \epsilon}. \label{zt0fin}
\eea
The $\beta$-functions evaluated from the eqs.~(\ref{zu0fin}), (\ref{zm0fin}) and
(\ref{zt0fin}) are written as follows
\bea
\beta_u &=& u^R[-\epsilon + {3u^R \over 2}+4t^R], \label{betau0} \\
\beta_m &=& m^R[-\epsilon + {3u^R \over 4} + m^R], \label{betam0} \\
\beta_t &=& t^R[-\epsilon -{u^R \over 8}+4m^R -4t^R]. \label{betat0}
\eea

The fixed points (FPs) are evaluated by equating the $\beta$-functions to zero. 
Noting that all of $u^R,m^R,t^R>0$ on physical ground, we discard any FP 
with negative values of the renormalized coupling constants. 
The positive semi-definite FPs are given by 
\begin{itemize}
\item FPI: Gaussian FP - $u^R=0,\,m^R=0,\,t^R=0$.
\item FPII: DP FP - $u^R=\frac{2\epsilon}{3},\,m^R=0,\,t^R=0$.
\item FPIII: $u_R=0,\,m^R=\epsilon,\,t^R=0$.
\item FPIV: $u^R={2\epsilon \over 3},\,m^R={\epsilon \over 2},\,t^R=0$. 
\item FPV: $u^R=0,\, m^R=\epsilon,\, t^R={3\epsilon \over 4}$.
\end{itemize}

Note that FPIV and FPV involves $m^R>0$ and $(m^R>0,t^R>0)$ respectively. What this 
means physically is that there is back mutation involved from mutants B to species A 
in both the FPs as the parameters $m$ and $t$ involves the back mutation coefficients 
$\lambda_2$ and $\lambda_g$.
The stability of the FPs can be found by evaluating the eigenvalues $\Lambda$ of the
stability matrix corresponding to each FP. The eigenvalues are found to be 
\begin{itemize}
\item FPI (Gaussian FP): The eigenvalues are
$\Lambda=-\epsilon,-\epsilon,-\epsilon$. All the eigenvalues
are negative indicating that the gaussian FP is unstable in all directions.
\item FPII (DP FP): The eigenvalues are $\Lambda=\epsilon, {-\epsilon
\over 2},{-13\epsilon \over 12}$. Thus, it is stable only along the
$u^R$-axis and unstable in all the other directions.
\item FPIII The eigenvalues are $\Lambda=-\epsilon,\epsilon,3\epsilon$.
Thus, FPIII is unstable only along the $u^R$-direction.
\item FPIV: The eigenvalues are $\Lambda=\epsilon, {\epsilon \over 2},
{11\epsilon \over 12}$. Thus this FP is stable along all directions. This is an 
important observation considering the fact that it involves back mutation from 
species B to A.
\item FPV: The eigenvalues are $\Lambda=2\epsilon,\epsilon,-3\epsilon$.
Thus FPV is unstable aong the $t^R$-axis but stable in the other two
directions.
\end{itemize}

The Wilson's flow are used to determine the critical exponents corresponding
to the different FPs. They are evaluated as follows:
\bea
\zeta_\rho=\mu{\partial \over
\partial\mu}\ln{Z_\rho^{-1}}\,,\, \zeta_{\hat\rho}= \mu{\partial \over
\partial\mu}\ln{Z_{\hat\rho}^{-1}}\,,\, \zeta_{D_\rho}=\mu{\partial \over
\partial\mu} \ln{Z_{D_\rho}^{-1}}\,,\, \zeta_\tau=\mu{\partial \over
\partial\mu}\ln{Z_\tau^{-1}}-2.\label{flow} \eea
 The critical exponents are derived from the flow functions (\ref{flow}) as
\bea
\eta_\rho &=& \eta_{\hat\rho}=-\zeta_\rho, \label{etarho} \\
{1 \over \nu} &=& -\zeta_\tau, \label{nu} \\
z_\rho &=& 2+\zeta_{D_\rho}. \label{dynrho}
\eea
We obtain 
\begin{itemize}
\item FPII (DP FP): The exponents are $\eta_\rho=\eta_{\hat\rho}=-{\epsilon \over 12}$,
$\nu^{-1}=2-{\epsilon \over 4}$,
dynamic exponent $z_\rho=2-{\epsilon \over 12}$.
\item FPIV: $z=2-{\epsilon \over 12}$, $\eta_\rho=\eta_{\hat\rho}=-{\epsilon \over 12}$,
and $\nu^{-1}=2-{\epsilon \over 4}$, which is exactlty equal to the DP critical
exponents. Thus this FP behaves like a DP FP, and displays weak dynamic scaling
making it physically acceptable. Note that this FP involves back mutation from mutant B to 
species A. 
\end{itemize}

\begin{figure}[htb]
\includegraphics[height=6cm]{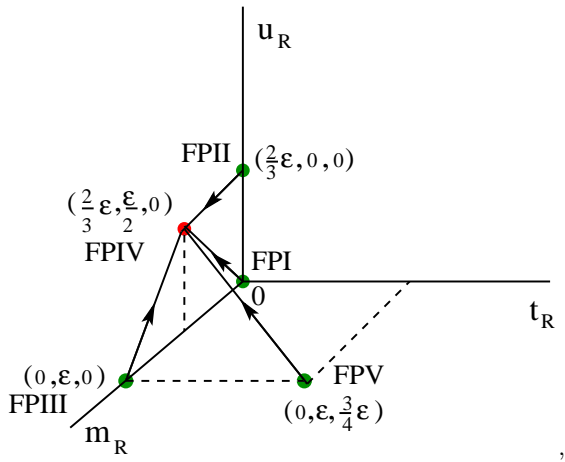},
\caption{(color online) A schematic DRG flow diagram showing the stable and 
unstable 
fixed points (FPs) of our model in the parameter space $u^R-m^R-t^R$. 
The stable FP is FPIV given by the red thick dot and the unstable 
FPs are given by thick green dots. All the FPs correspond to weak dynamic 
scaling. Note that the flow is from the unstable FPs to the 
stable FPIV.}
\label{fpdiag}
\end{figure}

Thus we find that FPII and FPIV are the physically acceptable FPs displaying
weak dynamic scaling as is expected in the $\theta^R\rightarrow 0$ limit (see 
below).
Also we see that ${\partial \beta_\theta \over \partial\theta^R}={u^R \over 8}>0$,
and ${\partial \beta_\theta \over \partial A^R}=0$, with $A=u,m,t$. This shows
that the weak dynamic scaling shown by the FP FPIV is stable along
the $\theta^R$ direction also. 
Surprisingly for FPIV, the critical exponents are all equal to the DP critical
exponents. One cannot say for sure if it falls under the DP universality class,
as we have calculated the exponents only upto one loop order. Higher loop
corrections are necessary to settle the issue. The DP FP or the FPII also shows
DP like critical exponents, which is not surprising given that the couplings
other than $u^R$ are taken to be zero. But this FP is unstable, 
which is due to the birth-death couplings of $\rho$ with $\phi$ in our model, 
unlike its 
analogue in the  DP model. Clearly, thus on stability grounds, we accept 
FPIV as the FP that describes the scaling of the AAPT in the present model. 
A schematic diagram of the fixed points in the parameter space is given in Fig.~\ref{fpdiag}. 

Finally, with the knowledge of 
$\eta_\rho,\eta_{\hat\rho}$, we now obtain the scaling of the renormalized 
equal-time correlator
\beq
\langle |\rho({\bf q},t)|^2\rangle\sim q^{2\eta_\rho}.
\eeq

The fact that the critical exponents at the FPIV are identical to those at 
the DP FP may be heuristically argued as follows. Notice that at FPIV, $t^R=0$, 
which means either $\psi^R$ or $\gamma^R$ or both of them are zero. Thus either 
$\lambda_g^R$ or $D_1^R$ or both are zero in this FP. If $\lambda_g^R=0$, the 
dynamics of $\rho$ is autonomous in the renormalized theory; consequently, 
DP-like exponents should be expected for the scaling properties of 
$\rho$-dynamics near AAPT. On the other hand, if $\lambda_G^R\neq 0$ but 
$D_1^R=0$, then the renormalized dynamics of $\phi$ is effectively noise free. 
Thus to the leading order, $\phi\approx\lambda_1^R/\lambda_2^R$ in the 
renormalised theory. If we substitute this in the renormlized version of 
(\ref{rho}), the effective renormalized equation for $\rho$ has the form of 
the basic Langevin equation for DP in terms of shifted coefficients. This 
indicates the DP-like exponents for $\rho$ near AAPT. This arugument is however 
only suggestive and cannot be used to claim the equality of the scaling 
exponents at FPIV with the same for the DP problem at higher order in 
perturbation theory. Due to the difference in the stability properties of 
FPII and FPIV already at the one-loop order, we tend to speculate that these 
two FPs will correspond to different values for the scaling exponents at 
higher loop orders. This can only be checked by rigorous higher order 
calculations which are beyond the scope of the present work.

\section{Correlations of $\phi$}

With the knowledge of the scaling at the AAPT, we can now 
calculate the renormalized correlator $C_{\phi\phi}$ of $\phi$. 
Notice that there are no anomalous dimension of $\phi$ and $\hat\phi$, 
nor there is any renormalization to $D_\phi$.  We linearize Eq.~(\ref{phifinal}) to obtain
\beq
 \frac{\partial\phi}{\partial t}= D_\phi\nabla^2 \phi + 
 (\lambda_3-\lambda_2\phi_0)\rho +\sqrt\rho\zeta.\label{phinew}
 \eeq
 Noting that $\langle\rho\rangle\rightarrow 0$ near AAPT, the scaling 
 of the renormalized correlation of $\phi$ may be obtained as
 \beq
 \langle |\phi({\bf q},\omega)|^2\rangle = \frac{1}{\omega^2 + D_\phi^2 q^4}
 {(\lambda_3-\lambda_2\phi_0)^2 \langle |\rho ({\bf q},\omega)|^2\rangle}.
 \eeq

 Now use the form  of renormalized $\langle |\phi({\bf q},\omega)|^2\rangle$ 
 at the stable FPs:
 \beq
 \langle |\rho({\bf q},\omega)|^2\rangle \sim \frac{q^{z+2\eta_\rho}}
 {\omega^2 + D_\rho q^{2z}},
 \eeq
 such that the equal-time correlator $\langle |\rho({\bf q},t)|^2\rangle
 \sim q^{2\eta_\rho}$. This yields
 \beq
 \langle|\phi({\bf q},t)|^2\rangle\sim\frac{1}{D_\phi}q^{2\eta_\rho-2-z}\exp(-D_\phi q^2 t),
 \eeq
revealing that (i) $z_\phi=2\neq z_\rho$ implying weak dynamic scaling 
and (ii) $\phi$ is spatially long-ranged correlated, since $2\eta_\rho-2-z <0$.

\section{Conclusion}
\label{con}

In this article we have proposed and studied a simple population dynamics 
model involving a species and its mutant:  species A (with density 
$\rho({\bf x},t)$) reproduces and dies, and also mutates to species B. We allow 
for a specific form of back-mutation from B to A that allows for an AAPT of A 
that we study here. In the absorbing state, the mutant B is conserved. We 
perform a one-loop perturbative DRG analysis
to extract the critical exponents of the AAPT. For reasons of analytical 
tractability, we analyze the model at low ($\theta_R\rightarrow 0$) 
Schmidt number. For $\theta_R\rightarrow 0$, 
we find weak dynamic scaling, i.e.,
$z_\rho<z_\phi=2$ at two FPs: FPII or the DP FP and FPIV, consistent with 
$\theta_R\rightarrow 0$. Interestingly, FPIV exhibits
scaling exponents which are same as those in the DP universality class or FPII. 
We believe this surprising feature is fortuitous as it is not likely to 
be preserved when higher order contributions are taken into account. In any 
case, FPII is unstable where as FPIV turns out to be stable in all the 
directions of the parameter space. 
In this article we have not attempted to obtain the FPs for finite $\theta_R$, 
for which strong dynamic scaling should follow, due to the algebraic 
complications involved. Nevertheless, from the overall stability of FPIV with 
$z_\rho<z_\phi$, together with $\beta_\theta >0$ at this FP and the fact that 
no stable FP is obtained for $\theta_R\rightarrow\infty$, we speculate that
 the AAPT in our model is indeed characterized 
by weak dynamic scaling only ($z_\rho <z_\phi$), precluding no stable FP for 
strong dynamic scaling. 
Further conclusive evidence of these will however require numerically obtaining 
the FPs from the zeros of the $\beta$-functions
(\ref{betau}-\ref{betapsi}). Interestingly, the scaling of $C_{\phi\phi}$ 
at the AAPT displays long-ranged spatial correlation, which is a consequence of 
the 
long-ranged $\rho$-fluctuations at the AAPT.


Our model has a highly simplified structure and is designed to study 
specific issues as discussed above. As a result, it lacks many details 
of a realistic population dynamics model. First of all, we have assumed an 
artificial form for back-mutation, which may be generalized and included in our 
model in a straight forward way.  In addition,
we have excluded the effects of
environment from our study, assuming a uniform surrounding in which the
interactions take place.  These may be included in our model in 
straightforward ways; see, e.g., Ref.~\cite{nil1}. In addition, the 
process of mutation is generally more complex than a simple conversion 
of one species to the other at a given fixed rate, as assumed here~\cite{mut}.
 Nonetheless, our model should be useful
 as a starting point to understand the critical behavior of 
 AAPT in generic population dynamics model with mutations or with multiple 
species. 
 We expect, our studies should be helpful in understanding the APPT 
 in rock-paper-scissors type systems \cite{erwin1,erwin2,erwin3,parker} where
one species  feeds on a second species B, which in turn feeds on another third
species that in turn feeds on the first species. We look forward to further 
theoretical studies along these lines.

\section{Acknowledgement}
The author would like to thank  A Basu for stimulating discussions, helpful suggestions 
and careful reviewing of the manuscript.

\appendix

\section{Bare vertex functions}
\label{appen1}

The bare vertex functions for the model can be found out by taking functional derivatives 
of the generating functional $\Gamma[\rho,\hat\rho,\phi,\hat\phi]$ with respect to the 
various fields in the model i.e., $\rho,\hat\rho,\phi,\hat\phi$. They are listed below:
\bea
&&\frac{\delta^2 \Gamma}{\delta\rho ({\bf k},\omega)\delta\hat\rho({\bf -k},-\omega)}=
 \Gamma_{\rho\hat\rho}=i\omega + D_\rho(-\tau + k^2),\\
 &&\frac{\delta^2 \Gamma}{\delta \phi ({\bf k},\omega)\delta\hat\phi({\bf -k},-\omega)}=
 \Gamma_{\phi\hat\phi}=(i\omega +D_\phi k^2), \\
  &&\frac{\delta^2 \Gamma}{\delta \hat\phi({\bf -k},-\omega)\delta\rho({\bf k},\omega)}=
 \Gamma_{\hat\phi\rho}=-\lambda_3,\\
&&\frac{\delta^3 \Gamma}{\delta \hat\rho({\bf q}_1,\omega_1)
\delta\rho ({\bf q}_2,\omega_2)\delta\rho({\bf -q}_1-{\bf
q}_2,-\omega_1 -\omega_2)}=\Gamma_{\hat
\rho\rho\rho}=\frac{D_\rho g_1}{2},\\
&&\frac{\delta^3 \Gamma}{\delta \hat\rho({\bf q}_1,\omega_1)
\delta\hat\rho ({\bf q}_2,\omega_2)\delta\rho({\bf -q}_1-{\bf
q}_2,-\omega_1 -\omega_2)}=\Gamma_{\hat
\rho\hat\rho\rho}=-\frac{D_\rho g_2}{2},\\
&&\frac{\delta^3 \Gamma}{\delta \hat\rho ({\bf k},\omega)\delta\rho
({\bf q},\Omega)\delta \phi ({\bf -k
-q},-\omega-\Omega)}=\Gamma_{\hat\rho\rho\phi}=-\lambda_g, \\
&&\frac{\delta^3 \Gamma}{\delta \hat\phi ({\bf k},\omega)\delta\phi
({\bf q},\Omega)\delta \rho ({\bf -k
-q},-\omega-\Omega)}=\Gamma_{\hat\phi\phi\rho}=\lambda_2, \\
&&\frac{\delta^3 \Gamma}{\delta \hat\phi ({\bf k},\omega)\delta\hat\phi
({\bf q},\Omega)\delta \rho ({\bf -k
-q},-\omega-\Omega)}=\Gamma_{\hat\phi\hat\phi\rho}=-D_1.
\eea

\section{Normalization conditions}
\label{appen2}
The renormalized vertex functions
when expressed in terms of the renormalized quantities can be written as follows:
\bea
\frac{\partial\Gamma_{\hat\rho\rho}}{\partial\omega}|_{({\bf
 k} =0,\omega=0)}&=&i,\\
 \frac{\partial\Gamma_{\hat\rho\rho}}{\partial k^2}|_{({\bf
 k}=0,\omega=0)} &=& D_\rho^R,\\
 \Gamma_{\hat\rho\rho}({\bf k}=0,\omega=0)&=&D_\rho^R\tau^R,\\
 \frac{\partial\Gamma_{\hat\phi\phi}}{\partial\omega}|_{({\bf
 k} =0,\omega=0)}&=&i,\\
\frac{\partial\Gamma_{\hat \phi\phi}}{\partial k^2}|_{({\bf
 k}=0,\omega=0)} &=& D_\phi^R,\\
 \Gamma_{\hat\phi\rho}({\bf k}=0,\omega=0)&=& -\lambda_3^R,\\
 \Gamma_{\hat\rho\rho\rho}({\bf k}=0,{\bf
 q}=0,\omega=0,\Omega=0)&=& \frac{D_\rho^Rg_1^R}{2},\\
 \Gamma_{\hat\rho\hat\rho\rho}({\bf k}=0,{\bf
 q}=0,\omega=0,\Omega=0)&=& -\frac{D_\rho^Rg_2^R}{2},\\
 \Gamma_{\hat\rho\rho\phi}({\bf k}=0,{\bf
 q}=0,\omega=0,\Omega=0)&=& -\lambda_g^R,\\
 \Gamma_{\hat\phi\phi\rho}({\bf k}=0,{\bf
 q}=0,\omega=0,\Omega=0)&=& \lambda_2^R,\\
 \Gamma_{\hat\phi\hat\phi\rho}({\bf k}=0,{\bf
 q}=0,\omega=0,\Omega=0)&=& -D_1^R.
 \label{renorm}
\eea

\end{document}